\documentclass[slac_one]{revtex4}
\usepackage{graphicx}
\usepackage{fancyhdr}
\begin{document}

\title{$W$ and $Z$ Boson Production at the ATLAS Experiment} 

%

\author{M. Schott \\on behalf of the ATLAS collaboration}
\affiliation{CERN, CH-1211, Geneve 23, Switzerland}

\begin{abstract}
The LHC experiments are close to collecting data and one of their first tasks is to test the electro-weak sector of the Standard Model. In this talk an overview of first physics measurements with events containing $W$ and $Z$ bosons is presented, such as the $W$ and $Z$ production cross-sections. Emphasis will be given to data-driven approaches used to extract trigger and lepton reconstruction efficiencies and to reject backgrounds.
\end{abstract}

\maketitle

\thispagestyle{fancy}


\section{Introduction} 

The study of the production of $W$ and $Z$ boson events at the LHC is fundamental in several respects. The calculation of higher order corrections to these simple, colour singlet final states is very advanced, with a residual theoretical uncertainty smaller than 1\%. Such precision makes $W$ and $Z$ production a stringent test of QCD. Moreover, a number of fundamental electroweak parameters can be accessed by high precision measurements and their transverse momentum distribution will provide more constraints on QCD. The ATLAS experiment (Ref. \cite{CiteATLAS}) aims to measure the overall production cross-section $\sigma$

\begin{equation}\label{eq:cross}
\sigma = \frac{N-B}{\epsilon \times A \times \int L dt}
\end{equation}

with an integrated luminosity of $50\,\mbox{pb}^{-1}$ already to high precision. Here, $N$ is the number of selected candidate events, $B$ is the number of background events, $\epsilon$ is the trigger and reconstruction efficiency\footnote{$\epsilon$ includes also the efficiency of the signal selection cuts.} and $A$ stands for the theoretical uncertainty on the detector acceptance.

It will be of special importance to determine the detector performance, via the precisely measured properties of the $Z$ boson, which we will discuss in section \ref{chap2}. The signal selection for $W$ and $Z$ boson events, both in electron and muon decay channel will be presented in section \ref{chap3} and \ref{chap4}. 

\section{\label{chap2}Detector Performance Determination with Data}

\subsection{Lepton Momentum Scale and Resolution}

The knowledge of the transverse momentum scale and resolution for high energetic leptons is essential for many physics analyses. The muons, resulting from a $Z$ boson decay, have mainly transverse momenta within a  range of $20$ GeV  and $60$ GeV. Hence,  one major approach for the determination of the scale and resolution within this $p_T$-range depends on a detailed study of the reconstructed $Z$ boson resonance. The $p_T$ resolution of the lepton reconstruction has a direct impact on the measured width, while the $p_T$ scale has a direct impact on the measured mean value.
 
In order to determine the lepton momentum scale and resolution, the following method is used. The $p_T$ resolution function predicted by Monte Carlo simulations is iteratively adjusted in its width and scale and the corresponding $Z$ boson mass distribution is calculated for each iterative step. This iterative process stops if the new predicted $Z$ boson mass distribution agrees within its statistical error to the measured distribution. We expect to determine the momentum scale for electrons and muons better than $1\%$, while the uncertainty on the resolution is smaller than $10\%$ for an integrated luminosity of $50\,\mbox{pb}^{-1}$.

\subsection{Lepton Trigger and Reconstruction Efficiency}

The trigger and reconstruction efficiencies for leptons can be determined in data via the so-called "tag and probe" method. The basic principle for the determination of the muon reconstruction and trigger efficiencies is based on the independence of the ATLAS Inner Detector and its Muon Spectrometer.
Two reconstructed, isolated and highly energetic Inner Detector tracks, with an invariant mass close to $91$ GeV are selected. If one of these tracks can be associated to a reconstructed muon track, it is called the ÔtagÕ muon. It ensures that one has selected a $Z\rightarrow\mu\mu$ event and one can probe with the second track the reconstruction or trigger efficiency. A systematic uncertainty of less than 1\% is expected for an integrated luminosity of $50\,\mbox{pb}^{-1}$.
 
The same principle can be applied for the electron trigger and reconstruction efficiency. In this case, this ÔprobeÕ electron is a reconstructed energy cluster in the calorimeter above $20$ GeV. In contrast to the muons, the background contribution for electrons cannot be neglected and must be determined via a background fit function. Also here, a systematic uncertainty of less than 1\% is expected for the chosen integrated luminosity.

\section{\label{chap3}$Z$ Boson Production}

A $20$ GeV single electron trigger has been chosen for the primary event selection for the electron decay channel. Two offline reconstructed and oppositely charged electrons must fulfill tight selection criteria, i.e. a good match of an electron signature and a reconstructed track. Moreover, both electrons must have transverse momenta above $20\,\mbox{GeV}$ and the resulting invariant mass must be between $80\,\mbox{GeV}$ and $100\,\mbox{GeV}$.

The signal selection in the muon decay channel is started by a $20\,$GeV single muon trigger. Two offline reconstructed and opposite charged tracks in the Muon Spectrometer must be matched to Inner Detector tracks and have also transverse momenta above $20\,\mbox{GeV}$. The invariant mass must be between $70\,\mbox{GeV}$ and $110\,\mbox{GeV}$. In contrast to the electron, we require that both muon tracks must appear isolated in the Inner Detector. 

The QCD and $W$ background contribution for both decay modes can be determined within data, while the contribution from the top pairs is estimated via Monte Carlo simulation. The theoretical uncertainties due to initial state radiation, parton shower model and PDFs have also been studied. An overview of the expected precision is shown in Table \ref{tableZee}.

The theoretical uncertainties on the acceptance are already dominating with an integrated luminosity of  $50\,\mbox{pb}^{-1}$. For larger integrated luminosities it is expected to constrain the uncertainties on PDFs via differential cross-section measurements and hence also to reduce the theoretical uncertainties for the overall cross-section measurement. The largest experimental uncertainty in this study is due to the determination of the lepton reconstruction and trigger efficiency in data.

\begin{table}[t]
\begin{center}
\caption{\label{tableZee}Expected results for the overall cross-section measurement with an integrated luminosity of $50\,\mbox{pb}^{-1}$. The 
uncertainty on N is statistical, the other errors are systematic. The systematic uncertainty combines effects due to background contribution, limited detector response knowledge (i.e. uncertainty on the efficiency $\epsilon$) and theoretical uncertainties on the acceptance. An overall luminosity uncertainty of $\delta L /L = 10\%$ should be counted in addition. }
\begin{tabular}{|l|c|c|c|c|c|c|}
\hline \textbf{Process} & N ($\times 10^4$)& B ($\times 10^4$) & $A\times\epsilon$ & $\delta A/A$ & $\delta \epsilon / \epsilon$ & $\sigma (\mbox{pb}) \pm (stat) \pm (sys)$ \\
\hline
$Z\rightarrow ee$ & $2.71\pm0.02$ &  $0.23\pm0.04$ & $0.246$ & $0.011$ & $0.03$ & $2016\pm16\pm72$ \\
$Z\rightarrow \mu \mu$ & $2.57\pm0.02$ & $0.010\pm0.002$ & $0.254$ & $0.011$ & $0.03$ & $2016\pm16\pm64$ \\
\hline
$W\rightarrow e\nu$ & $22.67\pm0.04$ & $0.61\pm0.92$ & $0.215$ & $0.023$ & $0.02$ & $20520\pm40\pm1060$ \\
$W\rightarrow \mu\nu$ & $30.04\pm0.05$ & $2.01\pm0.12$ & $0.273$ & $0.023$ & $0.02$ & $20530\pm40\pm630$ \\
\hline
\end{tabular}
\label{l2ea4-t1}
\end{center}
\end{table}

\section{\label{chap4}$W$ Boson Production}

\begin{figure}[tbp]
  \begin{minipage}[b]{8.3cm}
    \includegraphics[height=50mm]{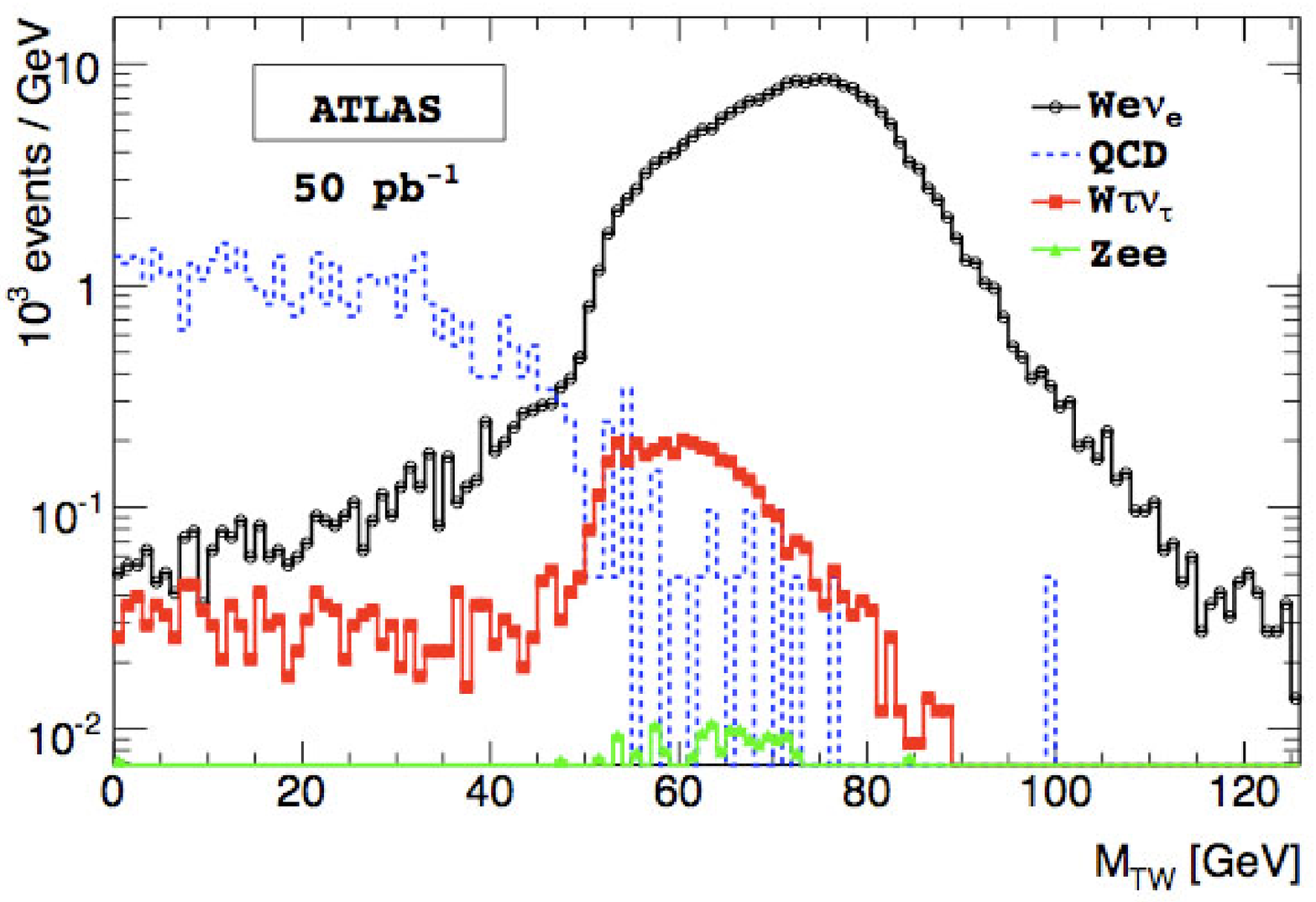}
  \caption{\label{labelWmassee}Transverse mass distribution of Monte Carlo simulated and selected events in the electron decay channel.}
  \end{minipage}
\hspace{0.5cm}
  \begin{minipage}[b]{8.3cm}
    \includegraphics[height=50mm]{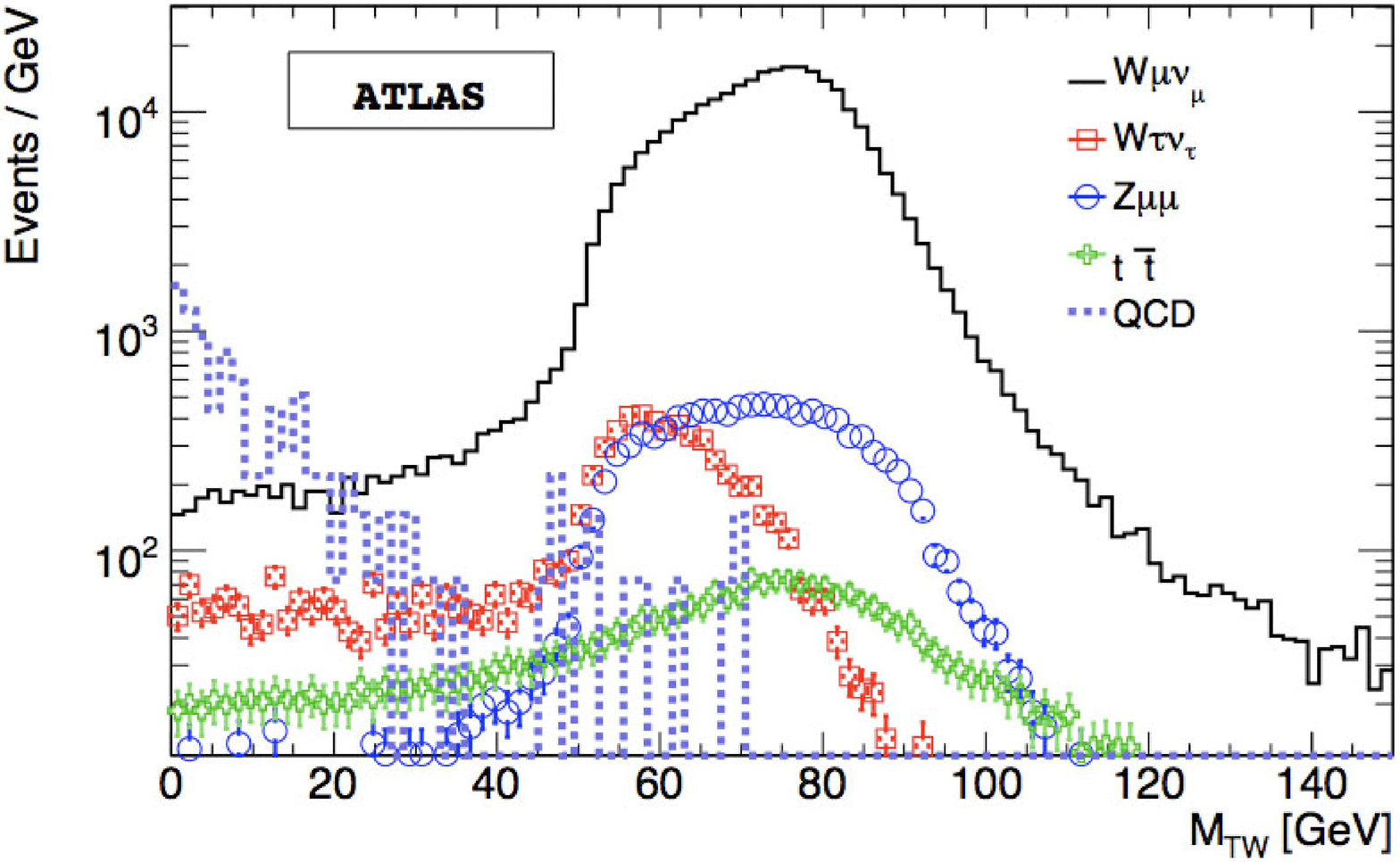}
  \caption{\label{labelWmassmumu}Transverse mass distribution of Monte Carlo simulated and selected events in the muon decay channel.}
  \end{minipage}

\end{figure}

The selection of $W\rightarrow l\nu$ events is based on the same triggered data as the corresponding $Z$ boson channels. All events are required to have a missing energy above $25\,$GeV. In the electron decay channel at least one reconstructed electron with a transverse momentum above $25\,$GeV is required. The resulting transverse mass must be above $40\,$GeV. In the muon decay channel we require at least one reconstructed and isolated muon with similar kinematic requirements. Also the transverse mass must be above $40\,$GeV. The expected transverse mass distribution after the applied cuts is shown in Figure \ref{labelWmassee} and Figure \ref{labelWmassmumu} for the electron and muon channel, respectively.

The largest background contribution in the muon channel comes from $Z$ boson decays, in which one muon leaves the detector unseen. This contribution can be estimated precisely from Monte Carlo simulations. An overview of the expected precision is also shown in Table \ref{tableZee}.

In contrast to the $Z$ boson case, we expect larger systematic uncertainties due to background contributions mainly from QCD. The background uncertainty is the dominating effect in the electron channel, while the theoretical uncertainty is still dominating for the muon decay. 

\section{Conclusion}

With an integrated luminosity of $50\,\mbox{pb}^{-1}$, the background and signal acceptance uncertainties contribute similarly to the measured cross-section uncertainty, at the level of 2-4\% depending on the channel neglecting the uncertainty on the integrated luminosity. Extrapolating to $1\,\mbox{fb}^{-1}$, all uncertainties are expected to scale with statistics, except the acceptance uncertainty which is theoretically limited. Without further studies of the differential cross sections to improve the theoretical understanding, the $W$ and $Z$ cross-sections cannot be measured to a precision better than about 2\%. A detailed description of these studies can be found in Ref. \cite{CiteCSCNote}.


\begin{thebibliography}{9}   

\bibitem{CiteATLAS}
The ATLAS Collaboration, G. Aad et al., 2008 JINST 3 S08003 

\bibitem{CiteCSCNote}
ATLAS Collaboration, Expected Performance of the ATLAS Experiment, Detector, Trigger and Physics,
CERN-OPEN-2008-020, Geneva, 2008, to appear. 


\end{thebibliography}
\end{document}